\documentclass{aipproc}

\usepackage{graphics}
\usepackage{color}

\usepackage{longtable}

\def\xx#1{\!\times\!10^{#1}}
\newcommand\pmax{p_{\mathrm{max}}}
\newcommand\rg{r_{\mathrm{g}}}
\newcommand\rgmax{r_{\mathrm{g,max}}}
\newcommand\dSNR{D_{\mathrm{snr}}}
\newcommand\EmisVol{V_{\mathrm{emis}}}
\newcommand\Vsk{V_{\mathrm{sk}}}
\newcommand\Rsk{R_{\mathrm{sk}}}
\newcommand\Emax{E_{\mathrm{max}}}
\newcommand\EmaxPro{E_{\mathrm{max,p}}}
\newcommand\EffRel{\epsilon_{\mathrm{rel}}}
\newcommand\DStemp{T_{\mathrm{p2}}}
\newcommand\DStp{T_{\mathrm{tp}}}
\newcommand\sun{{\odot}}
\newcommand\epRatio{(e/p)_{\mathrm{rel}}}
\newcommand\TempPro{T_{p0}}
\newcommand\TempRatio{T_{\mathrm{e2}}/T_{\mathrm{p2}}}
\newcommand\EnSN{E_{\mathrm{sn}}}
\newcommand\Mej{M_{\mathrm{ej}}}
\newcommand\pcc{cm$^{-3}$}
\newcommand\kmps{km s$^{-1}$}
\newcommand\muG{$\mu$G}
\newcommand\ProDenUpS{n_{p0}}
\newcommand\mpc{m_{\mathrm{p}} \, c}
\newcommand\brem{bremsstrahlung}

\newcommand\syn{synchrotron}

\newcommand\IC{inverse-Compton}
\newcommand\pion{pion-decay}
\newcommand\Rtot{r_\mathrm{tot}}
\newcommand\Rsub{r_\mathrm{sub}}
\newcommand\gameff{\gamma_{\rm eff}}
\newcommand\MAZ{M_\mathrm{A0}}
\newcommand\MSZ{M_\mathrm{S0}}
\newcommand\alf{Alfv\'en}
\newcommand\etainjP{\eta_\mathrm{inj,p}}
\newcommand\etainjE{\eta_\mathrm{inj,e}}
\newcommand\etamfp{\eta_\mathrm{mfp}}
\newcommand\TP{test-particle}
\newcommand\NL{nonlinear}

\newcommand\DSA{diffusive shock acceleration}
\newcommand\iec{i.e., }
\newcommand\egc{e.g., }
\newcommand\etal{et al.}
\newcount\Inum
\newcount\IInum
\newcount\IIInum
\newcount\IVnum
\def\I{\global\multiply\IInum by 0 \global\multiply\IIInum by 0
            \global\multiply\IVnum by 0 \global\advance \Inum by 1
            {\the\Inum. }}
\def\II{\global\multiply\IIInum by 0\global\multiply\IVnum by 0
       \global\advance \IInum by 1 {\the\Inum.\the\IInum. }}
\def\III{\global\multiply\IVnum by 0\global\advance \IIInum by 1
            {\the\Inum.\the\IInum.\the\IIInum. }}
\def\IV{\global\advance \IVnum by 1
            {\the\IVnum. }}
\newcount\reffnum
\def\refnumDCE{\global\advance \reffnum by 1 {\the\reffnum }}
\def\newrefnum{\reffnum=0}
\newcount\listno
\def\listDCE{\global\advance \listno by 1 {(\the\listno) }}

\newcount\listnorom
\def\listromanDCE{\global\advance \listnorom by 1 
{\lowercase\expandafter{\ (\romannumeral\listnorom)}\ }}
\def\newlistroman{\listnorom=0}

\begin{document}

\title{The Cosmic Ray -- X-ray Connection: Effects of Nonlinear Shock
Acceleration on Photon Production in SNRs}

\author{Donald C.~Ellison  
}
\affiliation{ 
Department of Physics, 
North Carolina State University,
 Box 8202, Raleigh NC 27695, U.S.A.;
\ don\_ellison@ncsu.edu 
%
}

\begin{abstract}
Cosmic-ray production in young supernova remnant (SNR) shocks is
expected to be efficient and strongly nonlinear. In nonlinear,
\DSA, compression ratios will be higher and
the shocked temperature lower than \TP , Rankine-Hugoniot
relations predict.  
Furthermore, the heating of the gas to X-ray emitting temperatures is
strongly coupled to the acceleration of cosmic-ray electrons and ions,
thus nonlinear processes which modify the shock, influence the
emission over the entire band from radio to gamma-rays and may have a
strong impact on X-ray line models.  Here we apply an algebraic model
of nonlinear acceleration, combined with SNR evolution, to model the
radio and X-ray continuum of Kepler's SNR.

\vskip12pt
{Proceedings of the ACE-2000 Symposium on {\it The
Acceleration and Transport of Energetic Particles Observed in the
Heliosphere},
January 5 - 8, 2000, Indian Wells, CA}

\end{abstract}

\maketitle

\vspace{-6 mm}
\section{Introduction}

More than twenty years of spacecraft observations in the heliosphere
have proven that collisionless shocks can accelerate particles with
high efficiency, \iec 10-50\% of the ram energy can go into superthermal
particles (\egc Eichler \cite{Eich81}; Gosling \etal\
\cite{GoslingEtal81}; 
Ellison \etal\ \cite{EMP90}; Terasawa \etal\ \cite{Terasawa99}). 
Energetic particles exist throughout the universe and shocks are
commonly associated with them, confirming that shock acceleration is
important beyond the heliosphere as well.  In fact, shocks in
supernova remnants (SNRs) are
believed to be the main source of Galactic cosmic rays, and 
these shocks 
are expected to be much stronger than those in
the heliosphere and can only be more efficient and nonlinear.

The conjecture that collisionless shocks are efficient accelerators is
strengthened by results from plasma simulations, which show efficient
shock acceleration consistent with spacecraft observations (\egc
Scholer, Trattner, \& Kucharek \cite{STK92}; Giacalone \etal\
\cite{GBSE97}), and other indirect evidence comes from 
radio emission from
SNRs (see Reynolds \& Ellison \cite{RE92}) and
equipartition arguments in AGNs and $\gamma$-ray bursts  (see
Blandford \& Eichler \cite{BE87} for an early review). 
There is also clear evidence that shocks can produce strong
self-generated turbulence. This has long been seen in heliospheric
shocks (\egc Lee \cite{Lee82, Lee83}; Kennel \etal\
\cite{KennelEtal84}; Baring \etal\ \cite{BaringEtal97}) and there is
evidence that it occurs at SNRs as well (\iec Achterberg, Blandford,
\& Reynolds \cite{ABR94}).

While the importance of nonlinear (NL) shock acceleration is evident,
NL solutions to \DSA\ are complicated and results are often unwieldy
and difficult to use for astrophysical applications. Therefore, we
have developed a simple, algebraic model of \DSA, based on more
complete studies, which includes the essential nonlinear effects
(Berezhko \& Ellison \cite{BEapj99}; Ellison, Berezhko, \& Baring
\cite{EBB2000}).  This technique is computationally fast and
easy-to-use, yet includes (i) the modification of particle spectra
when the backpressure from energetic ions smooths the shock structure,
and (ii) the influences on the shock dynamics when the magnetic
turbulence is strongly amplified by wave-particle interactions.

The complications of NL shock acceleration and the many parameters
required to characterize it are offset somewhat by the fact that the
entire particle distribution function, from thermal to the highest
energies, is interconnected and must be 
accounted for self-consistently with a nonthermal tail
connecting the quasi-thermal population to the energetic one.  
Because energy is conserved, a change in
the production efficiency of the highest energy particles {\it must}
impact the thermal properties of the shock heated gas and vice
versa. If more energy goes into relativistic particles, less is
available to heat the gas. In contrast, the power laws assumed by
test-particle models have no connection with the thermal gas, energy
conservation does not constrain the normalization of the power law,
and the spectral index can be changed with no feedback on the thermal
plasma.  Furthermore, there is a direct linkage between protons and
electrons (which produce most of the photon emission associated with
shocked gas) in nonlinear models, so the entire emission from
radio to gamma-rays, plus cosmic-ray observations, can, in principle,
be used to constrain the models.

Here we describe some of the nonlinear features expected to occur in
young SNRs and investigate some implications of efficient cosmic-ray
production on the broad-band continuum from
Kepler's SNR. We refer to Berezhko \& Ellison \cite{BEapj99} and
Ellison, Berezhko, \& Baring \cite{EBB2000} for details of the NL
shock model and its application to particle and photon production in
SNRs.
Work on the NL X-ray line emission from Kepler is in progress, i.e.,
Decourchelle, Ellison, \& Ballet \cite{DEB2000}.  Previous
test-particle calculations of the X-ray emission from Kepler have been
reported by Borkowski, Sarazin, \& Blondin \cite{BSB94}, who used a
two-dimensional hydrodynamic simulation, and by Rothenflug \etal\
\cite{RMCB94}, who investigated the emission from the reverse shock.
The work of Decourchelle \& Ballet \cite{Decour94} is mentioned below.

\vspace{-6 mm}
\section{Nonlinear Shock Acceleration}

The nonlinear effects in shock acceleration are of two basic kinds:
(i) the self-generation of magnetic turbulence by accelerated
particles and (ii) the modification (\iec smoothing) of the shock
structure by the backpressure of accelerated particles.
Briefly, (i) occurs when counter-streaming accelerated particles
produce turbulence in the upstream magnetic field which amplifies as
it is convected through the shock. This amplified turbulence results
in stronger scattering of the particles, and hence to more
acceleration, quickly leading to saturated turbulence levels near
$\delta B/B \sim 1$ in strong shocks.  The wave-particle interactions
produce heating in the shock precursor which may be observable.
Effect (ii) results in the overall compression ratio, $\Rtot$, being
an ever increasing function of Mach number (as if the effective ratio
of specific heats $\gameff \to 1$ and analogous to radiative shocks),
\iec 
\begin{equation}
\label{eq:machsonic}
\Rtot \simeq 1.3 \, \MSZ^{3/4} \quad {\rm if} \quad M_{S0}^2>M_{A0}
\ , 
\end{equation}
or by
\begin{equation}
\label{eq:machalf}
\Rtot \simeq 1.5\, M_{A0}^{3/8}
\end{equation}
in the opposite case ($\MSZ$ is the
sonic and $\MAZ$ is the \alf\ Mach number). Simultaneously, shock
smoothing causes the viscous subshock to be weak ($\Rsub \ll \Rtot$)
and the temperature of the shocked gas to drop below the
\TP\ value.  
\begin{figure}
\resizebox{\columnwidth}{!}{\includegraphics{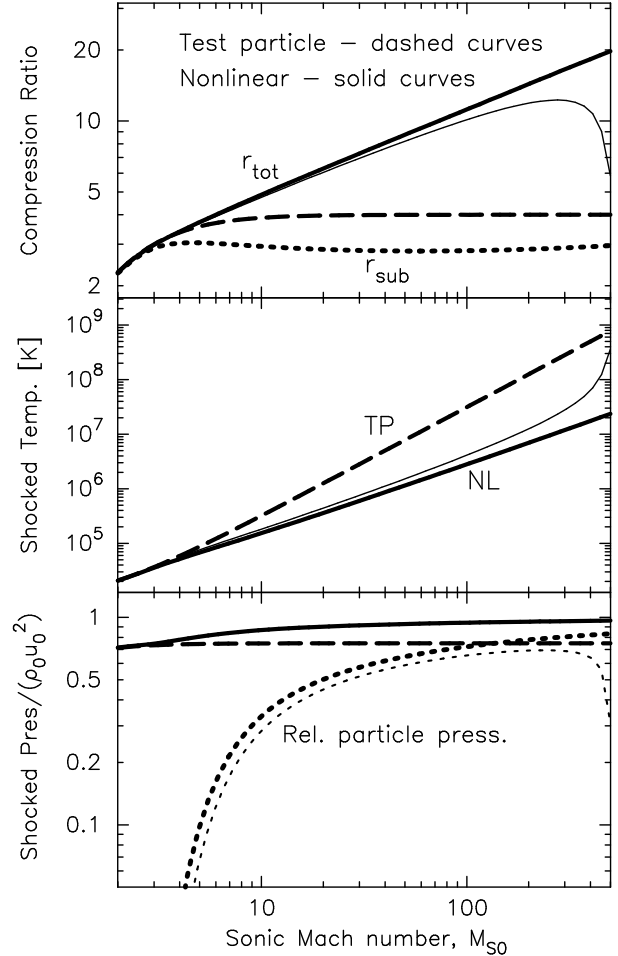}}
\caption{
Comparison of \NL\ and \TP\ shock results as a function of upstream
sonic Mach number. The top two panels show that, at $\MSZ > 100$, the
downstream density can be $>5$ times {\it larger} and the shocked
temperature $>10$ times {\it smaller} if the acceleration is NL. The
bottom panel shows that for $\MSZ > 100$ and $\etainjP=10^{-3}$, more
than 70\% of the pressure in the shocked gas ends up in relativistic
particles in NL shock acceleration.  These results depend on
parameters other than $\MSZ$, and in particular are a strong function
of $\etainjP$, particularly at high $\MSZ$. The heavy-weight curves
are for $\etainjP= 10^{-3}$ and the light-weight curves are for
$\etainjP= 10^{-4}$.}
\label{fig:tp_nl}
\end{figure}

There are many parameters associated with NL shock acceleration (see
Ellison, Berezhko, \& Baring \cite{EBB2000} for a listing), but the
most important ones that determine the solution are the Mach numbers
(\iec the shock speed, pre-shock density, and magnetic field) and the
injection efficiency, $\etainjP$ (\iec the fraction of total protons
which end up with superthermal energies).  As described in Berezhko \&
Ellison (\cite{BEapj99}), we use \alf\ heating in the precursor which
reduces the efficiency compared to adiabatic heating.  Significantly,
parameters typical of young SNRs should result in NL acceleration.

Figure~\ref{fig:tp_nl} shows a comparison between \TP\ (dashed lines)
and \NL\ results as a function of sonic Mach number, $\MSZ$.  As
illustrated in the top panel, the overall compression ratio, $\Rtot$,
is generally $>4$ and can increase without limit in the NL case for
$\etainjP = 10^{-3}$. Simultaneously, the subshock compression ratio,
$\Rsub$, is less than 4 (typically $\Rsub \sim 3$).  The overall
compression determines the shocked density, while the subshock is
mainly responsible for heating the gas.  Thus, the shocked gas will be
cooler (middle panel) and denser if the acceleration is NL compared to
a \TP\ (TP) case. This is an important consideration for X-ray line
models.  The particle spectrum at the highest energies is determined
by $\Rtot$, and thus will be flatter than in a TP shock. However, the
spectral shape at lower energies depends on $\Rsub$, indicating that
the particle spectrum will go from being steeper than the TP
prediction at low energies to flatter at the highest energies, this is
the so-called concave signature of NL shock acceleration.  The bottom
panel shows that a large fraction of the shocked pressure can end up
in relativistic particles (dotted line) for large $\MSZ$.  These
results depend on the injection efficiency assumed and the turnover in
$\Rtot$ at large $\MSZ$ seen in the $\etainjP=10^{-4}$ example is a
transition to a strong, {\it unmodified} shock (see Berezhko \&
Ellison \cite{BEapj99} for a full discussion).

\begin{figure}
\resizebox{\columnwidth}{!}{\includegraphics{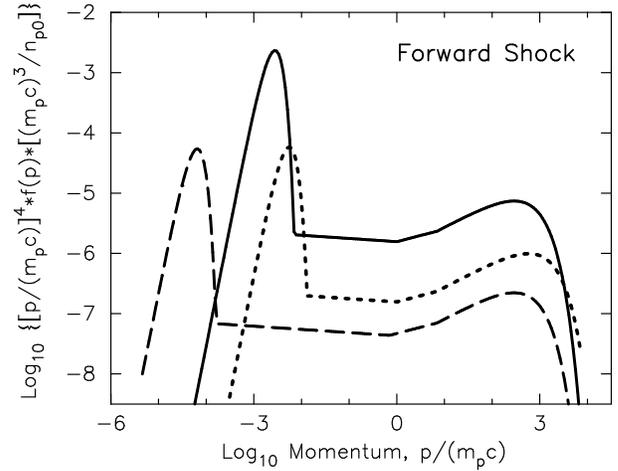}}
\caption{
Downstream phase space distribution functions, $f$, versus momentum,
$p$. We have multiplied $f(p)$ by $[p/(\mpc)]^4$ to flatten the
spectra, and by $[(\mpc)^3/\ProDenUpS]$ to make them dimensionless
($\ProDenUpS$ is the upstream proton number density).
The solid line shows protons, the dotted line shows helium, and the
dashed line shows electrons.
}
\label{fig:fp_kepler}
\end{figure}

\begin{figure}
\resizebox{\columnwidth}{!}{\includegraphics{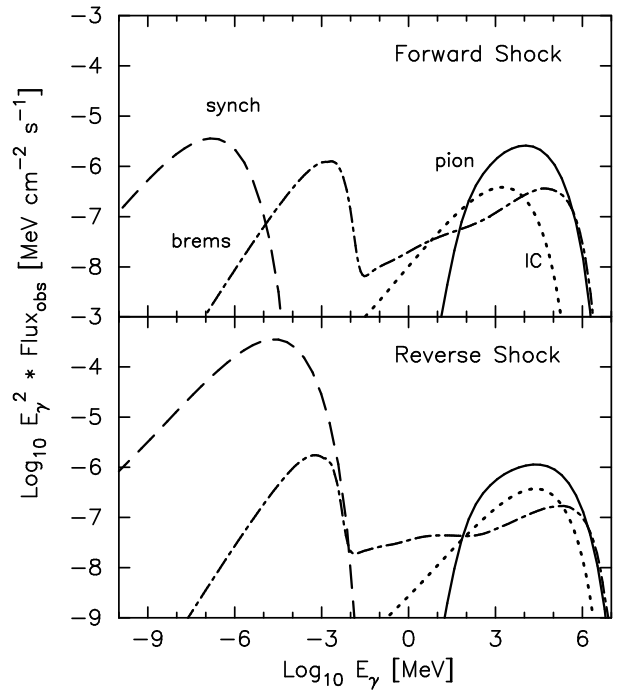}}
\caption{
Photon spectra from the forward (FS) and reverse shocks (RS)
obtained with parameters used to generate the spectra shown in
Figure~\ref{fig:fp_kepler}. Notice that emission in the X-ray band
(1-100 keV) is totally \brem\ for the FS but has a \syn\ component in
the RS.
Emission at gamma-ray energies is dominated by the FS, while the radio
comes totally from the RS.
}
\label{fig:photon_kepler}
\end{figure}

In Figure~\ref{fig:fp_kepler} we show the phase space distribution
functions, $f(p)$, for electrons (dashed line), protons (solid line),
and helium (dotted line). The parameters used for this forward shock
are those discussed below for Kepler's SNR and given in
Table~\ref{tab:KeplerTable}.
These particles produce photons from \syn, \brem, \IC, and \pion\
processes (see Baring \etal\ \cite{BaringEtal99} for details) and
these spectra are shown in Figure~\ref{fig:photon_kepler} for both the
forward and reverse shocks.  As discussed below, we have chosen
parameters to cause the reverse shock to contribute substantially to
the total X-ray emission, but this choice impacts the emission over
the entire range from radio to gamma-rays. While the radio \syn\ comes
primarily from the reverse shock, the forward shock dominates emission
at TeV energies with a \pion\ component. Any changes in parameters to
accommodate the X-ray observations must be consistent with
observational constraints in all other bands.


\vspace{-6 mm}
\section{Application to Kepler's SNR}

We now apply our model to the radio and X-ray emission seen in
Kepler's SNR. The X-ray emission shows the presence of lines and these
have been interpreted, in models without acceleration (see
Decourchelle \& Petre \cite{Decour99}; Decourchelle
\etal. \cite{Decour2000}), as coming from the ejecta material heated
by the reverse shock. However, these previous models also required the
presence of a power law photon continuum, presumably coming from \syn\
emitting TeV electrons (\egc Reynolds \cite{Reynolds96}).

\begin{figure}
\resizebox{\columnwidth}{!}{\includegraphics{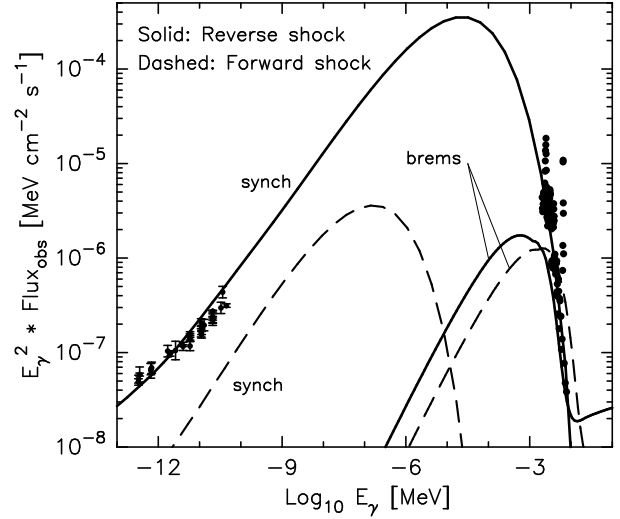}}
\caption{
Radio and X-ray data for Kepler's SNR compared to the emission from
the forward and reverse shocks.  The radio data is from Reynolds \&
Ellison \cite{RE92} and X-ray data is adapted from Decourchelle \&
Petre \cite{Decour99} (Decourchelle, private communication).  We use a
standard $\nu \, F_{\nu}$ representation and have suppressed the error
bars on the X-ray points.  
}
\label{fig:kepler}
\end{figure}

Here, we use the evolutionary model of Truelove \& McKee
\cite{Truelove99} to obtain the forward and reverse shock parameters
at the $\sim 395$ yr age of Kepler's SNR and, using these shock
parameters, calculate the NL particle and then photon spectra.
Figure~\ref{fig:kepler} shows the \syn\ and \brem\ emission from
Figure~\ref{fig:photon_kepler} and indicates that we are able to
obtain a good fit to the data within the constraint that the continuum
emission from the {\it reverse shock} contribute substantially to the
X-ray emission (our model only produces continuum which lies below any
X-ray emission lines).
Since the forward and reverse shocks have very different emission
profiles (Figure~\ref{fig:photon_kepler}), the ability to resolve the
individual shocks in radio and X-rays will be exceptionally
important for constraining the models.

We note that the relation between the line and continuum emission is
not straightforward and requires a detailed calculation as is
currently in progress, i.e., Decourchelle, Ellison, \& Ballet
\cite{DEB2000}. This is particularly the case in young SNRs because,
while the lines and continuum both depend on the electronic
temperature in the shocked gas, the lines also depend strongly on the
ejecta composition and on the ionization state. The ionization state
can be strongly influenced by non-equilibrium effects and by the
history of the shocked plasma.
What can be said, however, is that the presence of lines restricts the
\syn\ intensity to some level comparable to (with a factor of 10 say)
or below the \brem\ continuum.

As mentioned above, NL shocks are extremely complicated with many
(often poorly defined) parameters and our purpose in this preliminary
work is only to give some indication of the issues involved when NL
effects are considered.  The basic point is that, if nonlinear
cosmic-ray production occurs, the X-ray modeling
cannot be done without considering emission in other bands,
particularly radio and gamma-rays (if available).
For Kepler we know that strong lines are present, indicating that the
thermal gas behind the reverse shock is contributing substantially to
the emission.  This strongly constrains the range of acceptable
parameters.  Previous TP models have also indicated that a continuum
component is likely to be present. If so, this is likely to be \syn\
from TeV electrons and must be consistent with the radio observations.

\newlistroman

The parameters used to generate Figure~\ref{fig:kepler} are given in
Table~\ref{tab:KeplerTable}. All terms in this Table are explained in
Ellison, Berezhko, \& Baring (\cite{EBB2000}) but some additional
explanation is required since we include here, for the first time, a
calculation of acceleration at the reverse shock.
\listromanDCE
The unshocked proton number density, $\ProDenUpS$, is
arbitrarily chosen for the FS and determined for the RS with
eqs.~(20), (28), and (30) in Truelove \& McKee \cite{Truelove99}.
\listromanDCE
The unshocked magnetic field, $B_0$, is  arbitrarily chosen and the
shocked field (which produces the \syn\ emission) is taken to be
$B_2 = \Rtot \, B_0$.
We allow for an independent $B_0$ for the reverse shock
and, in fact, the only way the RS can contribute substantially to the
X-ray emission is for the RS $B_0$ to be much larger (25 vs. 1.5 \muG)
than the FS $B_0$.
\listromanDCE
The supernova energy, $\EnSN$, and ejecta mass, $\Mej$, are standard
values. 
\listromanDCE
We assume the same unshocked proton temperature, $\TempPro = 10^4$ K,
for the FS and RS. The solutions are relatively insensitive to this
parameter.
\listromanDCE
The shocked electron to proton temperature ratio, $\TempRatio$, is a
free parameter in the NL shock model  and is set to 1 for both shocks.
\listromanDCE
The injection efficiency, $\etainjP$, is an important model parameter
determining the overall acceleration efficiency. Values much larger
than $\etainjP = 2\xx{-4}$ tend to produce strong nonthermal tails on the
\brem\ emission which may be inconsistent with the observations. Values much
smaller than $2\xx{-4}$ yield \TP\ solutions with typical temperatures
higher than allowed by the X-ray observations. We note that
adjustments in $\etainjP$ produce extremely large modifications in the
TeV results and can be highly constrained with gamma-ray observations.
\listromanDCE
The electron to proton ratio at relativistic energies, $\epRatio$, is
an arbitrary parameter in the simple model but is expected to be
between 0.01 and 0.05 if  $\epRatio$ in galactic cosmic rays is
typical of that produced by the strong shocks in young SNRs.  This
factor is very important for determining the \pion\ contribution to
TeV gamma-rays. Furthermore, lowering $\epRatio$ lowers the overall
emission in the radio to X-ray band but increases the relative
contribution of thermal to \syn\ photons in the X-ray band.  A firm
determination of the \syn\ contribution to X-rays, combined with a
gamma-ray detection, will help determine $\epRatio$, one of the most
important unknown parameters in shock acceleration.
\listromanDCE
The maximum energy cosmic rays obtain in the NL model depends on the
scattering mean free path, $\lambda$, which is assumed to be,
\begin{equation}
\label{eq:mfp}
\lambda = 
\etamfp \, \rgmax \, \left ( \rg / \rgmax\right )^\alpha
\ ,
\end{equation}
where $\etamfp$ is taken to be independent of particle momentum
(Baring \etal\ \cite{BaringEtal99}, use the notation $\eta$), $\rg=
p/(q B)$ is the gyroradius in SI units, $\rgmax$ is the gyroradius at
the maximum momentum, $\pmax$, and $\alpha$ is a constant parameter
($\alpha = 1 = \etamfp$ is roughly the Bohm limit). Here we only
consider $\alpha=1$.  A fairly large value of $\etamfp=15$ (giving a
maximum cosmic-ray energy well below that obtained in the Bohm limit)
is required to avoid having the X-ray emission totally dominated by
\syn\ from TeV electrons.
\listromanDCE
We assume that the initial density profile in the ejecta has a power
law density distribution, $\rho \propto r^{-n}$, with $n=9$, and that
the unshocked ISM is uniform.

The output values are also given in Table~\ref{tab:KeplerTable}.  Of
these, $\Vsk$ and $\Rsk$ are taken directly from the Truelove \& McKee
\cite{Truelove99}) solution.  Once these are obtained, the Mach
numbers are determined (given the relevant input parameters), and then
the NL shock model determines the compression ratios, temperatures,
and particle spectra. Using the particle spectra, the continuum
emission is calculated.  It is important to notice that the shocks are
highly efficient and NL.  They yield total compression ratios $ > 4$,
shocked temperatures nearly 10 times lower than the corresponding \TP\
shocks, and place the majority of the total energy flux, $\EffRel$,
into relativistic particles (mainly protons),
yielding $\gameff$'s less than 5/3.

Finally, our model includes a rough estimate of the emission volume
\begin{equation}
\label{eq:Emission}
\EmisVol \approx (4\pi/3) \,\Rsk^3 / \Rtot
\ ,
\end{equation}
as described in Ellison, Berezhko, \& Baring \cite{EBB2000}. This volume is
considerably less than the total remnant volume.  If we assume a
distance to Kepler of $\dSNR = 5$ kpc (\egc Decourchelle \& Ballet
\cite{Decour94}), the normalization given by eq.~(\ref{eq:Emission}),
and shown in the Table, matches the observations.

\begin{table}
\begin{tabular}{cccc}
\hline
Input parameters &Forward shock & Reverse shock \\
\hline
$\ProDenUpS$ [\pcc]
&0.5
&1.25 
\\
$B_0$ [\muG]
&1.5
&25
\\
$\EnSN$ [10$^{51}$ erg]
&1 
&--- 
\\
$\Mej$ [$M_{\sun}$] 
&5 
&---
\\
$\TempPro$ [K] 
&$10^4$ 
&$10^4$ 
\\
$\TempRatio$ 
&1 
&1 
\\
$\etainjP$
&$2\xx{-4}$ 
&$2\xx{-4}$
\\
$\epRatio$ 
&0.03 
&0.03
\\
$\etamfp$ 
&15 
&15 
\\
$n$ 
&9 
&9
\\
\hline
Output values & & \\
\hline
$\Vsk$ [\kmps] 
&4100 
&1750
\\ 
$\Rsk$ [pc] 
&2.5 
&2.1 \\ 
$\MSZ$ 
&260 
&110 
\\ 
$\MAZ$ 
&$1040$ 
&43 \\ 
$\Rtot$ 
&11
&6.4 \\ 
$\Rsub$ 
&3.9 
&3.5 \\
$B_2$ [\muG] 
&16 
&160 
\\
$\EmaxPro$ [eV]\tablenote{This is the maximum proton energy, but since
\syn\ losses are unimportant here, electrons have the same $\Emax$.}
&$4.1\xx{11}$ 
&$1.7\xx{12}$ \\ 
$\etainjE$ \tablenote{This is the electron injection efficiency.}
&$5.1\xx{-4}$ 
&$6.3\xx{-4}$
\\ 
$\DStemp$ [K] 
&$2.8\xx{7}$ 
&$1.3\xx{7}$ \\ 
$\DStp$ [K]\tablenote{This is the temperature the shock gas would
have obtained if no acceleration took place.}
&$2.0\xx{8}$ 
&$9.4\xx{7}$ \\ 
$\EffRel$ 
&0.79 
&0.54 \\
$\gameff$ 
&1.40
&1.45 \\
\hline
Flux parameters & & \\
\hline
$\dSNR$ [kpc]
&5
&---
\\
$\EmisVol$ [pc$^3$]
&6.0
&6.4
\\
\hline
\end{tabular}
\caption{Parameters for Kepler's SNR model}
\label{tab:KeplerTable}
\end{table}

\vspace{-6 mm}
\section{Discussion and Conclusions}

X-ray line and continuum emission contains a vast amount of
information on supernova (SN) ejecta elemental composition, ISM
density and elemental composition, the SN explosion energy, and the
mass of ejecta.
In addition to heating the plasma, the forward and reverse shocks
accelerate some fraction of the shocked material to cosmic-ray
energies and this acceleration is believed to be quite efficient,
removing energy from the thermal plasma (\egc Kang \& Jones
\cite{KJ91}; Dorfi \& B\"ohringer \cite{DorfiB93}; Berezhko,
Ksenofontov, \& Petukhov \cite{BereKP99}). Despite the expected
efficiency of shock acceleration, virtually all current X-ray line
models assume that the shocks that heat the gas {\it do not} place a
significant fraction of their energy in cosmic rays (exceptions to
this are Chevalier \cite{Chev83} and Dorfi \cite{Dorfi94}).
Here, we investigated the broad-band continuum emission expected in
Kepler's SNR from efficient shock acceleration, by coupling
self-similar hydrodynamics (Chevalier \cite{Chev83}; Truelove \& McKee
\cite{Truelove99}) with nonlinear diffusive shock acceleration (\egc
Berezhko, Ksenofontov, \& Petukhov \cite{BereKP99}).
We were able to show that the radio and X-ray continuum can be fit
with reasonable parameters in a way that allows the reverse shock to
contribute substantially to the total X-ray emission. This constraint
is required since X-ray line models of Kepler (\egc Decourchelle \&
Petre \cite{Decour99}) require emission from the shock-heated,
metal-rich ejecta material.

This preliminary calculation is not fully self-consistent for several
reasons. Most importantly, we use self-similar results (\iec Truelove
\& McKee \cite{Truelove99}) to model the SNR evolution.
These solutions neglect the effects of energetic particle escape from
the FS and assume that the ratio of cosmic-ray pressure to total
pressure at the shock front is a constant. They also assume that
$\gameff = 5/3$. As indicated in the bottom panel of
Figure~\ref{fig:tp_nl}, the total shocked pressure doesn't differ much
with or without cosmic-ray production and we have demonstrated (\iec
Decourchelle, Ellison, \& Ballet \cite{DEB2000}) that NL shock
results do not change dramatically over most of the age of Kepler for
typical values of the injection parameter, $\etainjP > 10^{-4}$.
However for lower $\etainjP$, the nonlinear solutions can have
test-particle, unmodified solutions at very high sonic Mach numbers
with a rapid transition to the NL solution as the Mach number drops
(see Fig.~\ref{fig:tp_nl} and
Berezhko \& Ellison \cite{BEapj99} for a detailed discussion).
The self-similar solutions we use are still approximate,
however, because we have not yet modified them for the change in
$\gameff$ that results when a substantial fraction of the shocked
pressure is in cosmic rays.
We also neglect cosmic-ray diffusion and assume they are spatially
coupled to the gas -- an excellent approximation for all but the
highest energy particles. However, the highest energy electrons
produce the X-ray \syn\ photons so there may be differences that are
not modeled in the emission volumes and other important parameters
between the radio and X-ray bands.
Finally, we have not included absorption in our models which is
probably not important for the Kepler radio observations, but will be
required to model the low energy X-rays.\footnote{Note that we have
only plotted the X-ray observations above 2 keV in
Figure~\ref{fig:kepler} to avoid conflict with the absorbed low energy
end of the X-ray distribution.}

Besides providing a more self-consistent model of photon production,
predictions from NL shock models provide a test of the fundamental
assumption that SNRs are the primary source of galactic cosmic-ray
ions. If this is so, the acceleration is almost certainly nonlinear
since 5-30\% of the total ejecta kinetic energy is required to
replenish cosmic rays as they escape from the galaxy.  Since shocks
put more energy into accelerated ions than electrons, nonlinear
effects seen in X-ray emission will be evidence for the efficient
shock acceleration of ions as well as electrons.  X-ray observations
potentially provide {\it in situ} information on cosmic-ray ion
production, complementing observations of pion-decay $\gamma$-rays in
this regard. Any inference of nonthermal tails on electron
distributions in X-ray observations will provide information on
electron injection, the least well understood aspect of shock
acceleration.  Such {\it in situ} information on high Mach number
shocks is available nowhere else.

Our modeling of Kepler's SNR suggests that typical source parameters
produce large nonlinear effects in the broad-band spectrum and suggest
that the test-particle approximations that are almost universally used
are inadequate for SNRs as young as this.
Besides the differences discussed above, one might expect that the
growth rate of the Rayleigh-Taylor instability will be greater in a
cosmic-ray modified shock because of the larger spatial gradients of
density, pressure, etc.  Furthermore, the high compression ratios
result in a considerably thinner region between the forward and
reverse shocks than predicted in the TP case (Decourchelle, Ellison,
\& Ballet \cite{DEB2000}). This places the contact discontinuity closer
to the shock and may make it easier for the Rayleigh-Taylor
``fingers'' to distort or overtake the FS, a situation that appears
difficult with normal TP parameters (Chevalier, Blondin, \& Emmering
\cite{CBE92}; Chevalier \& Blondin \cite{CB95}).
Another important difference concerns electron heating and
equilibration. The higher densities in the NL models mean that
electron heating is much more efficient than in TP shocks. Our initial
calculations \cite{DEB2000} suggest
that it may be possible to obtain full equipartition between electrons
and ions, at least for high values of $\etainjP$, in the shocked
ejecta.


\vspace{-6 mm}
\section{Acknowledgments}
I wish to thank the organizers of the ACE-2000 Symposium for putting
on a very useful and enjoyable meeting and for providing support. I'm
especially grateful to A. Decourchelle for helpful comments and
suggestions and to A. Decourchelle and L. Sauvageot for furnishing the
Kepler X-ray data.


\vspace{-6 mm}

\begin{thebibliography}{}

\newrefnum

\bibitem[\refnumDCE]{ABR94}
Achterberg, A., Blandford, R.D.,
Reynolds, S.P. 
\aa{94}{281}{220}

\bibitem[\refnumDCE]{BaringEtal99}
Baring, M.G., Ellison, D.C., Reynolds, S.P., Grenier, I.A.,
\& Goret, P.,
\apj{99}{513}{311}

\bibitem[\refnumDCE]{BaringEtal97}
Baring, M.G., Ogilvie, K.W., Ellison, D.C., \& Forsyth, R.J.,
\apj{97}{476}{889}

\bibitem[\refnumDCE]{BEapj99}
Berezhko, E.G., Ellison, D.C.,
\apj{99}{526}{385}

\bibitem[\refnumDCE]{BereKP99} 
Berezhko, E.~G., Ksenofontov, L., \& Petukhov, S.~I., 
\icrcsaltlake{4}{431}

\bibitem[\refnumDCE]{BE87}
Blandford, R.D., \&  Eichler, D., 
\phyrepts{87}{154}{1}

\bibitem[\refnumDCE]{BSB94}
Borkowski, K.J., Sarazin, C.L., \&  Blondin, J.M.,
\apj{94}{429}{710}

\bibitem[\refnumDCE]{CB95} 
Chevalier, R.A., \& Blondin, J.M.,
 \apj{95}{444}{312} 

\bibitem[\refnumDCE]{CBE92} 
Chevalier, R.A., Blondin, J.M., \& Emmering, R.T.,
 \apj{92}{392}{118} 

\bibitem[\refnumDCE]{Chev83} 
Chevalier, R.A. \apj{83}{272}{765} 

\bibitem[\refnumDCE]{Decour94}
Decourchelle, A., \& Ballet, J.,
\aa{94}{287}{206}

\bibitem[\refnumDCE]{DEB2000}
Decourchelle, A., Ellison, D.C., \& Ballet, J., 
in preparation.

\bibitem[\refnumDCE]{Decour99}
Decourchelle, A., \& Petre, R., 1999, Astron. Nachr., 320, 203 

\bibitem[\refnumDCE]{Decour2000}
Decourchelle, A., et al., 2000, in preparation 

\bibitem[\refnumDCE]{Dorfi94} 
Dorfi, E. A.
\apjs{94}{90}{841}

\bibitem[\refnumDCE]{DorfiB93}
Dorfi, E.A., \& B\"ohringer, H.
\aa{93}{273}{251}

\bibitem[\refnumDCE]{Eich81}
Eichler, D. \apj{81}{247}{1089}

\bibitem[\refnumDCE]{EBB2000}
Ellison, D.C., Berezhko, E.G., \& Baring, M.G.,
2000, \apjpress - {\bf astro-ph/0003188}

\bibitem[\refnumDCE]{EMP90}
Ellison, D.C., M\"obius, E., \& Paschmann, G.,
\apj{90}{352}{376}

\bibitem[\refnumDCE]{GBSE97}
Giacalone, J., Burgess, D., Schwartz, S.J., Ellison, D.C., \&
Bennett, L. 
\jgr{97}{102}{19,789} 

\bibitem[\refnumDCE]{GoslingEtal81}
Gosling, J.T., Asbridge, J.R., Bame, S.J., Feldman, W.C., Zwickl,
R.D., Paschmann, G., Sckopke, N., and Hynds, R.J.
\jgr{81}{86}{547}

\bibitem[\refnumDCE]{KJ91}
Kang, H. \& Jones, T.~W. \mnras{91}{249}{439}

\bibitem[\refnumDCE]{KennelEtal84}
Kennel, C.F., Edmiston, J.P., Scarf, F.L., Coroniti, F.V.,
Russell, C.T., Smith, E.J., Tsurutani, B.T., Scudder, J.D., Feldman,
W.C., Anderson, R.R., Moser, F.S., and Temerin, M.,
\jgr{84}{89}{5436}

\bibitem[\refnumDCE]{Lee82}
Lee, M.A.,
\jgr{82}{87}{5063}
                                             
\bibitem[\refnumDCE]{Lee83}
Lee, M.A., 
\jgr{83}{88}{6109}

\bibitem[\refnumDCE]{Reynolds96} 
Reynolds, S.P. 
\apjlet{96}{459}{L13} 

\bibitem[\refnumDCE]{RE92}
Reynolds, S.P., \& Ellison, D.C.  \apjlet{92}{399}{L75}

\bibitem[\refnumDCE]{RMCB94}
Rothenflug, R., Magne, B., Chieze, J.P., \&
Ballet, J.,
\aa{94}{291}{271}

\bibitem[\refnumDCE]{STK92}
Scholer, M., Trattner, K.J., \& Kucharek, H.
\apj{92}{395}{675}

\bibitem[\refnumDCE]{Terasawa99} 
Terasawa, T., \etal\ 
\icrcsaltlake{6}{528} 

\bibitem[\refnumDCE]{Truelove99}
Truelove, J.K., \& McKee, C.F.:
\apjs{99}{120}{299}

\end{thebibliography}

\def\itt{\rm }
\def\bff{\rm }
\def\aa#1#2#3{ 19#1, {\itt A.A.,} {\bff #2}, #3}
\def\aasup#1#2#3{ 19#1, {\itt A.A. Suppl.,} {\bff #2}, #3}
\def\aj#1#2#3{ 19#1, {\itt A.J.,} {\bff #2}, #3}
\def\anngeophys#1#2#3{ 19#1, {\itt Ann. Geophys.,} {\bff #2}, #3}
\def\anngeophysic#1#2#3{ 19#1, {\itt Ann. Geophysicae,} {\bff #2}, #3}
\def\annrev#1#2#3{ 19#1, {\itt Ann. Rev. Astr. Ap.,} {\bff #2}, #3}
\def\apj#1#2#3{ 19#1, {\itt Ap.J.,} {\bff #2}, #3}
\def\apjlet#1#2#3{ 19#1, {\itt Ap.J.(Letts),} {\bff  #2}, #3}
\def\apjpress{{\itt Ap. J.,} in press}
\def\apjletpress{{\itt Ap. J.(Letts),} in press}
\def\apjs#1#2#3{ 19#1, {\itt Ap.J.Suppl.,} {\bff #2}, #3}
\def\apjsub#1{ 20#1, {\itt Ap.J.}, submitted.}
\def\app#1#2#3{ 19#1, {\itt Astroparticle Phys.,} {\bff #2}, #3}
\def\astrolets#1#2#3{ 19#1, {\itt Astronomy Letters.,} {\bff #2}, #3}
\def\asr#1#2#3{ 19#1, {\itt Adv. Space Res.,} {\bff #2}, #3}
\def\araa#1#2#3{ 19#1, {\itt Ann. Rev. Astr. Astrophys.,} {\bff #2},
   #3}
\def\araapress{{\itt Ann. Rev. Astr. Astrophys.,} in press.}
\def\ass#1#2#3{ 19#1, {\itt Astr. Sp. Sci.,} {\bff #2}, #3}
\def\eos#1#2#3{ 19#1, {\itt EOS,} {\bff #2}, #3}
\def\icrcplovdiv#1#2{ 1977, in {\itt Proc. 15th ICRC(Plovdiv)},
   {\bff #1}, #2.}
\def\icrcparis#1#2{ 1981, in {\itt Proc. 17th ICRC(Paris)},
   {\bff #1}, #2.}
\def\icrcbang#1#2{ 1983, in {\itt Proc. 18th ICRC(Bangalore)},
   {\bff #1}, #2.}
\def\icrclajolla#1#2{ 1985, in {\itt Proc. 19th ICRC(La Jolla)},
   {\bff #1}, #2.}
\def\icrcmoscow#1#2{ 1987, in {\itt Proc. 20th ICRC(Moscow)},
   {\bff #1}, #2.}
\def\icrcadel#1#2{ 1990, in {\itt Proc. 21st ICRC(Adelaide)},
   {\bff #1}, #2.}
\def\icrcdub#1#2{ 1991, in {\itt Proc. 22nd ICRC(Dublin)},
  {\bff #1}, #2.}
\def\icrccalgary#1#2{ 1993, in {\itt Proc. 23rd ICRC(Calgary)},
  {\bff #1}, #2.}
\def\icrcrome#1#2{ 1995, in {\itt Proc. 24th ICRC(Rome)},
  {\bff #1}, #2.}
\def\icrcromepress{ 1995, {\itt Proc. 24th ICRC(Rome)}, in press.}
\def\icrcdurban#1#2{ 1997, {\itt Proc. 25th Int. Cosmic Ray Conf.
    (Durban),} {\bff #1}, #2.}
\def\icrcsaltlake#1#2{ 1999, {\itt Proc. 26th Int. Cosmic Ray Conf.
    (Salt Lake City),} {\bff #1}, #2.}
\def\icrcsaltlakedate#1#2#3{ 19#1, {\itt Proc. 26th Int. Cosmic Ray Conf.
    (Salt Lake City),} {\bff #2}, #3.}
\def\icrcsaltlakepress#1#2{ 19#1, {\itt Proc. 26th Int. Cosmic Ray Conf.
    (Salt Lake City),} paper #2.}
\def\grl#1#2#3{ 19#1, {\itt G.R.L., } {\bff #2}, #3}
\def\jcp#1#2#3{ 19#1, {\itt J. Comput. Phys., } {\bff #2}, #3}
\def\jetp#1#2#3{ 19#1, {\itt JETP, } {\bff #2}, #3}
\def\JETP#1#2#3{ 19#1, {\itt JETP, } {\bff #2}, #3}
\def\JETPlet#1#2#3{ 19#1, {\itt JETP Lett., } {\bff #2}, #3}
\def\jgr#1#2#3{ 19#1, {\itt J.G.R., } {\bff #2}, #3}
\def\jpG#1#2#3{ 19#1, {\itt J. Phys. G: Nucl. Part. Phys., } {\bff #2}, #3}
\def\mnras#1#2#3{ 19#1, {\itt M.N.R.A.S.,} {\bff #2}, #3}
\def\nature#1#2#3{ 19#1, {\itt Nature,} {\bff #2}, #3}
\def\nucphys#1#2#3{ 19#1, {\itt Nuclear Phys. B,} {\bff #2}, #3}
\def\pss#1#2#3{ 19#1, {\itt Planet. Sp. Sci.,} {\bff #2}, #3}
\def\pf#1#2#3{ 19#1, {\itt Phys. Fluids,} {\bff #2}, #3}
\def\phyrepts#1#2#3{ 19#1, {\itt Phys. Repts.,} {\bff #2}, #3}
\def\pr#1#2#3{ 19#1, {\itt Phys. Rev.,} {\bff #2}, #3}
\def\prD#1#2#3{ 19#1, {\itt Phys. Rev. D,} {\bff #2}, #3}
\def\prl#1#2#3{ 19#1, {\itt Phys. Rev. Letts,} {\bff #2}, #3}
\def\pasj#1#2#3{ 19#1, {\itt Pub. Astro. Soc. Japan,} {\bff #2}, #3}
\def\pasp#1#2#3{ 19#1, {\itt Pub. Astro. Soc. Pac.,} {\bff #2}, #3}
\def\revgeospphy#1#2#3{ 19#1, {\itt Rev. Geophys and Sp. Phys.,}
   {\bff #2}, #3}
\def\rgsp#1#2#3{ 19#1, {\itt Rev. Geophys and Sp. Phys.,}
   {\bff #2}, #3}
\def\rmp#1#2#3{ 19#1, {\itt Rev. Mod. Phys.,} {\bff #2}, #3}
\def\rpp#1#2#3{ 19#1, {\itt Rep. Prog. Phys.,} {\bff #2}, #3}
\def\science#1#2#3{ 19#1, {\itt Science,} {\bff #2}, #3}
\def\sp#1#2#3{ 19#1, {\itt Solar Phys.,} {\bff #2}, #3}
\def\spu#1#2#3{ 19#1, {\itt Sov. Phys. Usp.,} {\bff #2}, #3}
\def\ssr#1#2#3{ 19#1, {\itt Space Sci. Rev.,} {\bff #2}, #3}

\end{document}